\documentclass{aa}  
\usepackage{graphicx}
\usepackage{txfonts}

\begin{document}

\title{Search for galactic Pevatron candidates in a population of unidentified $\gamma$-ray sources}

   \author{Gerrit Spengler}

   \institute{Institut f\"ur Physik, Humboldt-Universit\"at zu Berlin, Newtonstr.
15, 12489 Berlin, Germany\\
\email{spengler@physik.hu-berlin.de}}

   \date{Received September 4, 2019; accepted December 10, 2019}

 
  \abstract
   {}
   {A list of Pevatron candidates is presented to enable deeper observations and dedicated analyses.} 
   {Lower limits on the energy cutoff for unidentified $\gamma$-ray sources detected in the HESS galactic plane survey are derived. Additional public data from 
   the VERITAS, HESS and Milagro experiments are used for MGRO J1908+06 to confirm the limit derived from the HESS galactic plane survey data and to enable further conclusions on the 
   presence of spectral breaks.}
   {Five Pevatron candidates are identified in the HESS galactic plane survey. The cutoff of the $\gamma$-ray spectrum for these sources is larger than 20 TeV at 90\% 
   confidence level. The $\gamma$-ray sources MGRO J1908+06 and HESS J1641-463, found to be Pevatron candidates in the analysis of the HESS galactic plane survey catalog, 
   had been discussed as Pevatron candidates before. For MGRO J1908+06, the lower limit on the $\gamma$-ray energy cutoff is $30$ TeV at $90\%$ confidence level. This is a factor of almost two 
   larger than previous results. Additionally, a break in the $\gamma$-ray spectrum at energies between $1$ TeV and $10$ TeV with an index change $\Delta\Gamma>0.5$ can be 
   excluded at $90\%$ confidence level for MGRO J1908+06.\\
   The energy cutoff of accelerated particles is larger than $100$ TeV at $90\%$ confidence level in a hadronic scenario for all five 
   Pevatron candidates. A hadronic scenario is plausible for at least three of the Pevatron candidates, based on the presence of nearby molecular clouds and supernova remnants.}
   {}


   \maketitle

\section{Introduction}
\label{introduction}
Theoretical models for the origin of cosmic rays (CRs) consider young supernova remnants (SNRs) as Pevatrons, 
i.e. as hadron accelerators for energies up to the "knee" of the cosmic-ray spectrum measured at around $3$ PeV \citep{baade_zwicky, hillas}. 
The detection of a neutral pion decay signature in $\gamma$-ray data from SNR observations confirmed that SNRs accelerate hadronic particles (see e.g. \cite{neutral_pion}, 
\cite{jogler_funk}). However, the maximal energy to which SNRs can accelerate hadrons is not determined and it is in particular unclear 
whether SNRs are sources of PeV CRs.\\
The association of a $\gamma$-ray source with a Pevatron can be tested with a measurement of the $\gamma$-ray spectrum above TeV energies. An exponential cutoff $E_\mathrm{cut,\;\gamma}$ 
in the $\gamma$-ray spectrum at a few $100$ TeV is expected when CRs are accelerated up to PeV energies and produce $\gamma$-rays in interactions with ambient material \citep{gabici}. 
The Pevatron nature of a $\gamma$-ray source is in turn constrained when an exponential cutoff at an energy well below $100$ TeV is detected in the observed $\gamma$-ray spectrum of the source.
For example, the recent detection of 
an exponential cutoff at $E_\mathrm{cut,\;\gamma}=3.5$ TeV in the $\gamma$-ray spectrum of Cassiopeia A rules out simple Pevatron models for this young galactic SNR \citep{casA}.\\
The most constraining lower limits on the energy cutoff $E_\mathrm{cut,\;h}$ of galactic hadron 
accelerators are currently derived from the diffuse $\gamma$-ray emission in the vicinity of the radio source SgrA* and for the $\gamma$-ray source HESS J1641-463. 
Assuming that hadronic particles generate the $\gamma$-ray emission from these sources, 
$E_\mathrm{cut,\;h}>600$ TeV and $E_\mathrm{cut,\;h}>100$ TeV is inferred at $90\%$ and $99\%$ confidence level (CL) respectively \citep{gc, igor}. 
More than $100$ isolated sources of TeV $\gamma$-rays are currently known. However, no systematic search for Pevatron candidates among the known $\gamma$-ray sources has been presented yet.\\
This paper is organized as follows. Section \ref{search_strategy} outlines the general Pevatron search method followed in this work. 
The later analysis relies in large parts on the public catalog of the HESS galactic plane survey (HGPS). This catalog is introduced in Sec. \ref{all_data}. A method to constrain energy cutoffs in 
measurements of powerlaw energy spectra is discussed in Sec. \ref{statistics_section}. The method is argued to be applicable to the HGPS catalog introduced before. 
A search for Pevatron candidates in the HGPS catalog is summarized in Sec. \ref{hess_survey_section}. For one Pevatron candidate, identified in the previous analysis of HGPS data, 
additional public data is available from the HESS, VERITAS and Milagro experiments. A combined analysis of these data is presented in Sec. \ref{j1908_section}. Finally, sections 
\ref{discussion_section} and \ref{conclusion_section} discuss the results and summarize the conclusions.

\section{Pevatron candidate search method}
\label{search_strategy}
The detection of TeV $\gamma$ rays from an isolated region of the sky can indicate cosmic sites where particles are accelerated to energies larger than $1$ TeV. 
However, the identification of the primary particle type is non-trivial when 
only TeV $\gamma$-ray data are available. A leptonic and a hadronic scenario are typically conceivable. Electrons are accelerated in the leptonic scenario and generate 
$\gamma$-rays via the inverse Compton up-scattering of low energy photons. This scenario is not relevant in the following search for hadronic Pevatrons as origins of galactic CRs. 
The unambiguous exclusion of the leptonic scenario for a given $\gamma$-ray source is, however, very challenging. For example, as recently discussed, 
even the production of $\gamma$-rays with an energy exceeding $100$ TeV is possible in a leptonic scenario \citep{highest_energy_gamma_ray}. The non-detection of a $\gamma$-ray energy 
cutoff at energies below few $100$ TeV is therefore not considered as a sufficient criteria for the identification of a hadronic Pevatron. 
Instead of an unambiguous identification of hadronic Pevatrons, the following search aims to find Pevatron candidates, 
i.e. $\gamma$-ray sources where the hadronic Pevatron scenario cannot be excluded. The resulting list of Pevatron candidates allows 
dedicated multi-messenger analyses with reduced statistical trial factors, involving e.g. searches for high energy neutrinos as fingerprints of hadronic interactions.\\
Many known galactic $\gamma$-ray sources are identified as pulsar wind nebulae (PWNe) for which the $\gamma$-ray emission is modeled in a leptonic scenario \citep{pwn}. Another large 
class of $\gamma$-ray sources is not yet identified with sources of radiation in other energy ranges. The search for new Pevatron candidates will be restricted to these 
unidentified $\gamma$-ray sources in the galactic disc. The motivation for the restriction to unidentified $\gamma$-ray sources is at least twofold. The exclusion of PWNe 
removes a large class of $\gamma$-ray sources for which the leptonic scenario is very likely. Additionally, \cite{snr_mc} discuss that the time period where SNRs act as Pevatrons might 
be short, i.e. less than $1000$ years. Easier than the direct detection of Pevatrons might therefore be the 
indirect identification via the detection of delayed $\gamma$-ray emission from molecular clouds in the vicinity of SNRs. \cite{snr_mc} discuss this scenario and motivate that 
unidentified $\gamma$-ray sources can be associated with molecular clouds illuminated by nearby SNRs.\\ 
Following \cite{hadronic_gamma_ray_cutoff}, the spectrum of $\gamma$
rays generated in the vicinity of a Pevatron in interactions of accelerated CRs
with ambient material is in the following modeled by
\begin{equation}
	\phi(E)= \phi_0 \left( \frac{E}{E_0} \right)^{-\Gamma}\exp(-\sqrt{16\lambda_\mathrm{h}E})\:\mathrm{.}
	\label{hadron_spectrum}
\end{equation}
Here, $\phi_0\exp(-\sqrt{16\lambda_\mathrm{h}E_0})$ is the flux normalization at energy $E_0$ 
and the parent CR spectrum has a cutoff energy of $E_\mathrm{cut,\;h}=1/\lambda_h$ \citep{hadronic_gamma_ray_cutoff}. 
Models for diffuse shock acceleration predict a powerlaw index $\Gamma\approx 2$ (see e.g. \cite{hinton_hofman}). Following this model, TeV $\gamma$-ray sources that are selected as 
Pevatron candidates must have a spectrum at TeV energies which is compatible with a powerlaw with index $\Gamma=2$. Additionally, the inferred lower limit on the energy cutoff $E_\mathrm{cut,\;h}$ in 
the hadronic model given by Eq. \ref{hadron_spectrum} must be at least $\mathrm{O}(100\;\mathrm{TeV})$, 
i.e. in the order of magnitude of the current best constraints derived for HESS J1641-463 and the $\gamma$-ray emission in the vicinity of SgrA*.\\
To further constrain a leptonic scenario for the generation of $\gamma$ rays, the presence of a break in the TeV $\gamma$-ray spectrum where the powerlaw index changes 
can be excluded. Consider, for example, a leptonic scenario in which inverse Compton losses dominate over synchrotron cooling.
A Klein-Nishina (KN) break is expected in the $\gamma$-ray spectrum at an energy $E_\mathrm{KN}$ which depends on the target photon field. For cosmic microwave background (CMB) target photons, 
the break is expected at few hundred TeV. For infrared target photons or optical starlight, the energy break is expected around $10$ TeV and $30$ GeV respectively 
\citep{hinton_hofman}. The $\gamma$-ray spectrum at energies far away from $E_\mathrm{KN}$ can be described by a broken powerlaw with 
$E_\mathrm{break}=E_\mathrm{KN}$
\begin{equation}
	\phi(E) =
    \begin{cases}
	    \phi_0   \left( \frac{E}{E_0} \right)^{-\Gamma} & E< E_\mathrm{break}\\
	    \phi_0  \left( \frac{E_\mathrm{break}}{E_0} \right)^{\Delta\Gamma}  \left( \frac{E}{E_0} \right)^{-\Gamma-\Delta\Gamma} & E > E_\mathrm{break}\;\mathrm{.}
    \end{cases}       
\label{broken_pl}
\end{equation}
It is expected that $\Delta\Gamma=\Gamma$ for the index change at the KN break (see e.g. \cite{hinton_hofman}).
The selection of sources with a powerlaw index around $2$ at TeV energies excludes leptonic scenarios where the inverse Compton target photons 
have a much larger energy than CMB photons, e.g. optical starlight, and the KN break must occur below TeV energies. However, leptonic scenarios with 
CMB target photons can typically not be constrained with TeV $\gamma$-ray data.\\
The selection of a $\gamma$-ray source as a Pevatron candidate in the following search is not sufficient to identify the source with a Pevatron. 
Neither can leptonic emission scenarios be ruled out nor are the lower limits on $E_\mathrm{cut,\;h}$ sufficient to argue for the acceleration of hadrons to PeV energies. 
Additionally, the selection of a $\gamma$-ray source as a Pevatron is not necessary for the association with a Pevatron. 
For example, the $\gamma$-ray source Cassiopeia A would not pass the selection due to the measured TeV cutoff discussed in Sec. \ref{introduction}. However, Cassiopeia A is still 
being discussed as a Pevatron candidate in multi-zone models \citep{cas_a_with_cutoff}.\\
Despite of the Pevatron candidate selection being neither necessary nor sufficient for the association of a $\gamma$-ray source with a Pevatron, 
selected sources can be a particularly interesting target for further observations. For example, targeted signal searches with instruments 
sensitive to high energy neutrinos towards few selected Pevatron candidates can be performed.

\section{HGPS data and quality selection}
\label{all_data}
The currently largest catalog of galactic TeV $\gamma$-ray sources is based on the HGPS. This dataset, 
as far as relevant for the later data analysis, and additional quality selection criteria are discussed in this section.

\subsection{HGPS data}
The HGPS comprises data acquired from 2004 to 2013 with the HESS I \citep{crab} array of Cherenkov telescopes in Namibia. 
Large parts of the Milky Way plane were observed to detect $\gamma$-ray sources in a nominal energy range of $0.2-100$ TeV with an angular resolution better than $0.1^\circ$. 
Observations of the Milky Way plane were performed in a scanning mode with individual observation runs of typically $28$ min towards different sky directions. 
In total, $78$ $\gamma$-ray sources were detected in the plane survey. 
The majority ($47$) of the detected sources are unidentified, i.e. no firm identification is currently possible with objects detected in other energy ranges.\\
The HGPS catalog is based on a general purpose analysis without special adoption for each individual source. Systematic effects of this analysis cannot be ruled out, 
in particular for sky regions with multiple $\gamma$-ray sources.\\
An adaptive ring background algorithm is used in the analysis of data from the HGPS. This algorithm 
estimates the number of signal ($N_\mathrm{ON}$) and background ($N_\mathrm{OFF}$) events and enables a calculation of the excess events 
$N_\gamma=N_\mathrm{ON}-\alpha N_\mathrm{OFF}$. The acceptance normalization factor $\alpha$ is obtained from independent observations of different sky regions without $\gamma$-ray source. 
Energy dependent information on $N_\gamma$, $N_\mathrm{ON}$, $N_\mathrm{OFF}$ and $\alpha$ is not part of the public HGPS catalog. However, 
spectral flux data for each $\gamma$-ray source detected in the HGPS is typically available in $6$ logarithmic energy bins. The bin centers $E_i$, $i=0,\dots,N-1$, of spectral points 
are related via $E_i=\xi^i E_\mathrm{min}$ with 
a binning factor $\xi$ where $E_\mathrm{min}$ is the center of the lowest energy bin. Statistical errors on the $\gamma$-ray flux are given as asymmetric 
$68\%$ confidence level (CL) intervals and originate from the Poisson counting error on $N_\mathrm{ON}$ and $N_\mathrm{OFF}$. Other error sources, e.g. an error on the acceptance normalization, 
are not included in the statistical error on the $\gamma$-ray flux.
A complete description of the HGPS dataset can be found in \cite{hess_plane}. 

\subsection{Data selection}
\label{data_selection}
A special quality selection is applied in this analysis of the public HGPS data. The energy bin-width for spectral points is required to be 
much larger than the instrumental energy resolution to avoid a statistical correlation between spectral points. 
Following \cite{crab}, the energy resolution of the HGPS analysis is assumed to be $15\%$ or better. Only sources in the HGPS with spectral energy binning factor $\xi>1.5$ are selected.\\
In the HGPS catalog, data on the error of spectral points is given as lower ($\sigma_\mathrm{low}$) and upper ($\sigma_\mathrm{high}$) error. 
The average error $\sigma=1/2\;(\sigma_\mathrm{low}+\sigma_\mathrm{high})$ is used in the analysis. Spectral points with an error asymmetry 
$\eta=\mathrm{max}(|\sigma-\sigma_\mathrm{low}|, |\sigma-\sigma_\mathrm{high}|)/\sigma>10\%$ are discarded from the analysis. 
Also discarded are spectral points $\phi$ which are detected at low 
significance level $S=\phi/\sigma<1.5$. Sources in the HGPS catalog are not analyzed when less than $5$ spectral points remain after the quality selection. 
In total $25$ out of $78$ sources are selected for further analysis. A summary of the selected HGPS data is given in Table \ref{source_car}.
\begin{table}
	\label{sel_hgps_data}
	\caption{Characterization of the HGPS spectral data for the $25$ selected $\gamma$-ray sources. The columns show the minimum, maximum and median of the distribution of the 
	quantity in the corresponding row for the selected sources. $E_\mathrm{min}$ and $E_\mathrm{max}$ are the minimal and maximal bin-center energy of selected spectral points. The livetime 
	is the dead-time corrected observation time as obtained from the HGPS dataset. $\Gamma$ and $\phi_0$ are the best fit index and flux normalization for a powerlaw model. 
	The binning factor $\xi$, error asymmetry $\eta$ and spectral point significance $S$ are defined in Sec. \ref{data_selection}.}
	\label{source_car}

	\centering
\begin{tabular}{llll}
	\hline\hline
                 Quantity &   Minimum &   Maximum &    Median \\
\hline
	$E_\mathrm{min}$ (TeV) &       0.3 &      1.47 &       0.4 \\
	$E_\mathrm{max}$ (TeV) &      6.32 &     31.62 &     12.71 \\
    $\xi$ &       1.6 &       2.5 &       2.4 \\
	     Livetime (h) &       4.0 &     259.0 &      17.0 \\
		 $\Gamma$ &      1.81 &      2.69 &      2.31 \\
	$\phi_0$ ($\mathrm{cm^{-2}s^{-1}TeV^{-1}}$) &  1.80e-13 &  1.56e-11 &  3.51e-12 \\
	$\eta$ (\%) &       0.0 &       8.8 &       1.7 \\
	S &       1.5 &      22.8 &       4.3 \\

	\hline
\end{tabular}
\end{table}

\section{Constraints on the energy cutoff of a powerlaw spectrum}
\label{statistics_section}
\begin{figure*}
	\centering
	\includegraphics[width=\hsize]{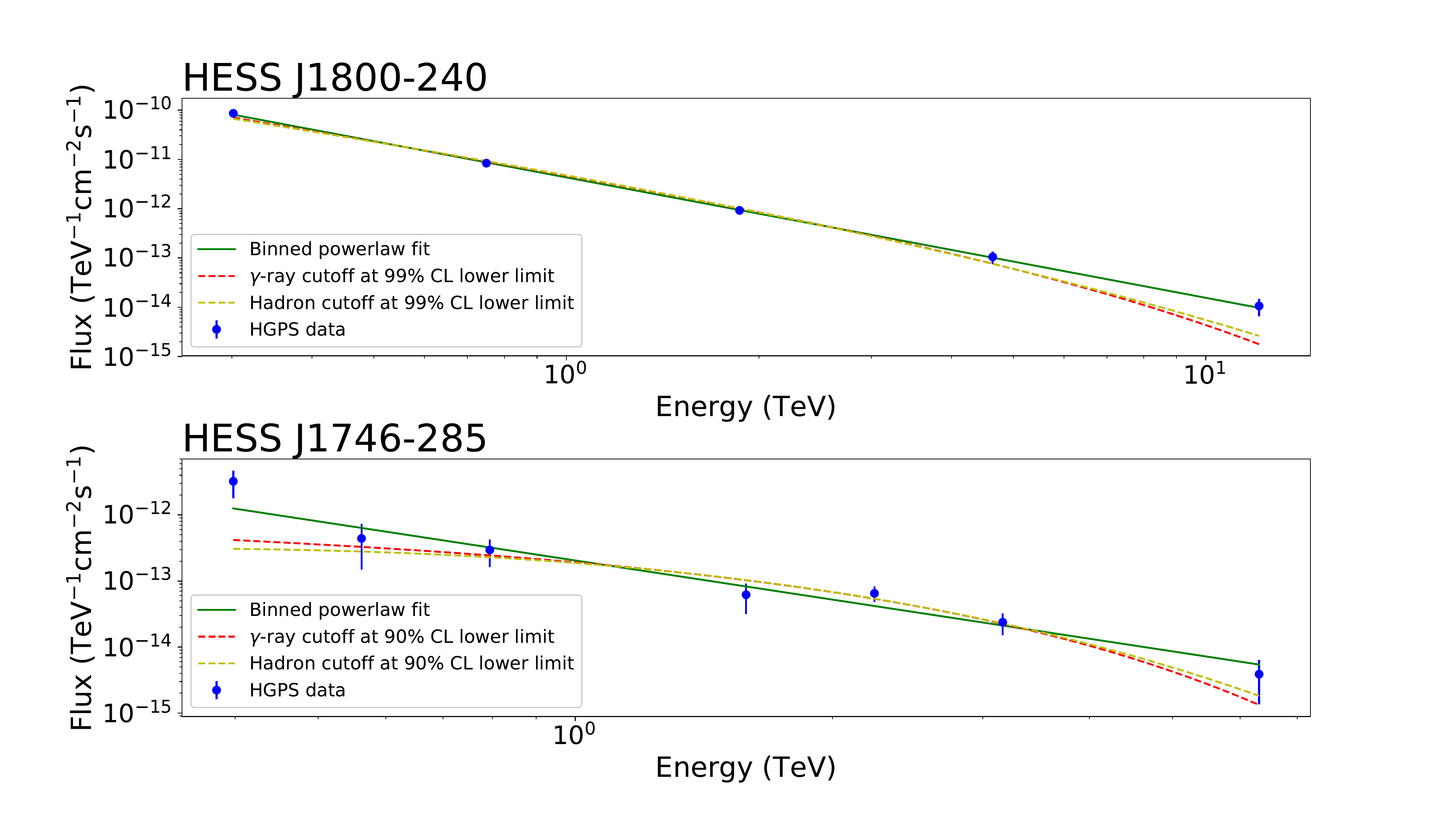}
	\caption{Spectral fits to the data from the HGPS source for which the most (upper panel) and least (lower panel) constraining lower limit on $E_\mathrm{cut,\;h}$ is derived. 
		 Shown in green are the best fits to the data with a powerlaw spectrum. 
		 Red lines are best fits to the data with a $\gamma$-ray powerlaw spectrum with exponential cutoff at the lower limit $E_\mathrm{cut,\;\gamma}$ derived from the HGPS data. 
		 Yellow lines are the corresponding best fits to the data in a hadronic model with cutoff at the lower limit $E_\mathrm{cut,\;h}$. Limits for the lower plot are at $90\%$ CL. 
		 In the upper panel, best fits with lower cutoff limit at $90\%$ CL are hardly distinguishable from the 
		 best powerlaw fit. $99\%$ CL best fits are therefore shown in this case.} 
	\label{best_and_worst_spectrum}
\end{figure*}

A statistical hypothesis test can be used to search for an exponential cutoff in a powerlaw spectrum. Via the inversion of the test, 
a confidence interval for the energy cutoff can be constructed. However, 
special care must be taken in the analysis of public data from the HGPS because the number of available spectral points per $\gamma$-ray source 
is typically small. The small sample size questions the application of frequently applied asymptotic results for the distribution of a test statistic when the null hypothesis is true. 
Additionally, systematic effects on the acceptance normalization cannot be ruled out and the catalog data is based on a general purpose analysis. 
Differences between the estimated and the true acceptance normalization due to instrumental effects 
could, for some observation runs, lead to outliers in the number of excess events which are not modeled with the published statistical error. 
Additional systematic effects can originate from the confusion of $\gamma$-ray sources in regions with complex morphology.\\
This section contains a description of statistical methods to constrain energy cutoffs in powerlaw spectra.
The application of the likelihood ratio (LR) test and the F-test for the presence of cutoffs in powerlaw spectra 
are discussed in section \ref{hypotheses}. In section \ref{robustness_section}, it is argued that the F-test is more robust against potential violations of 
the normal error model of the HGPS catalog data than the LR-test. A method to derive a lower limit on the energy cutoff of a powerlaw 
$\gamma$-ray energy spectrum is discussed in section \ref{confidence_interval_section}.
\subsection{Models and hypotheses tests}
\label{hypotheses}
\begin{figure*}
	\centering
	   \includegraphics[width=\hsize]{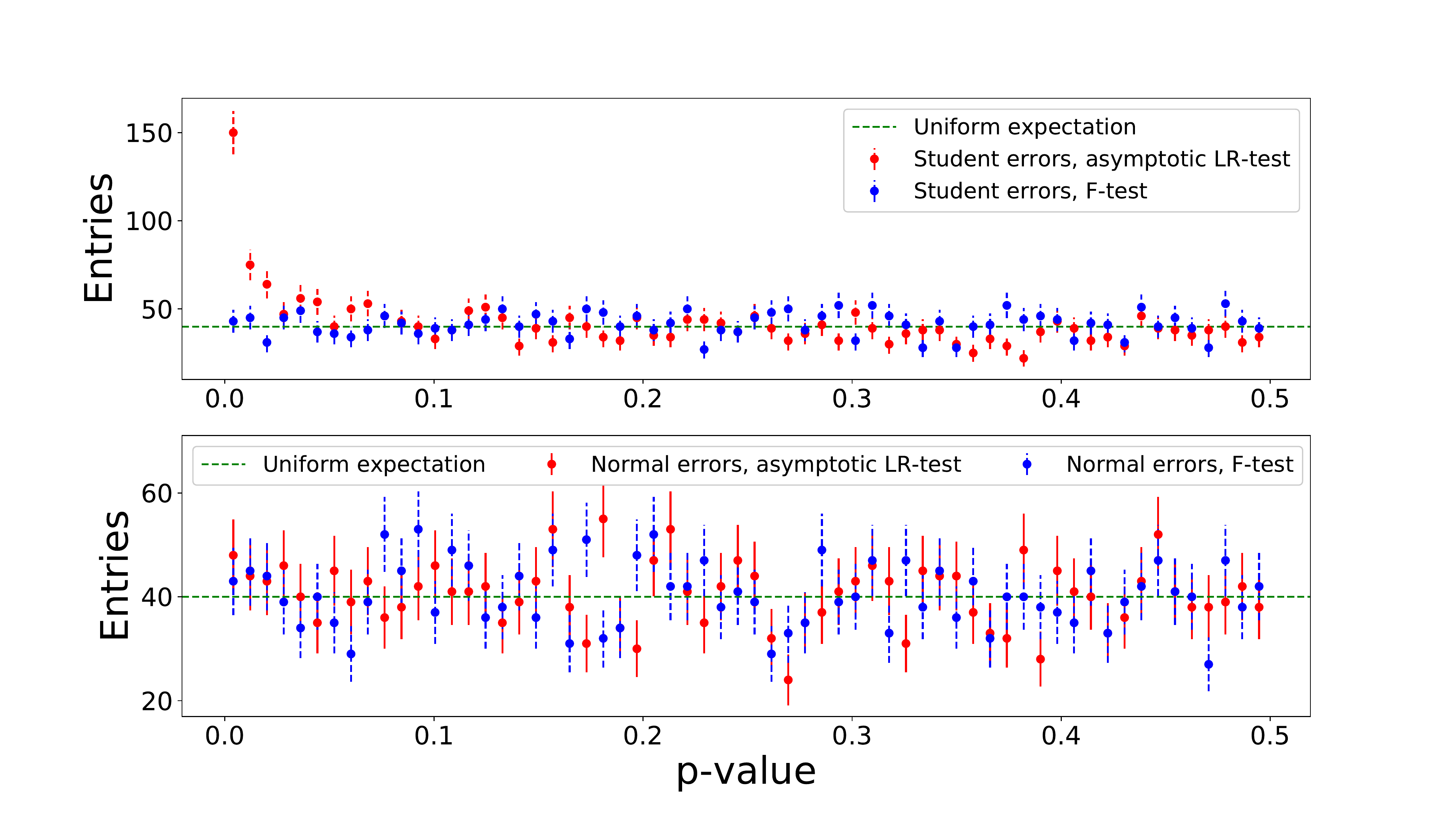}
	   \caption{Distribution of p-values obtained in LR- and F-tests in a Monte Carlo simulation of true null hypotheses. The expectation for the p-value distribution is uniform in the 
	   interval $0\leq p\leq 0.5$ where $50\%$ of the p-values are predicted. The other half of the p-values is, due to the constraint to positive energy cutoffs, expected at $p=1$ 
	   (see also appendix \ref{critical_value_appendix}).
	   The upper panel shows the resulting p-value distribution in the range $0\leq p \leq 0.5$ 
	   when the simulated spectral points disperse around the true data according to a Student t-distribution with $5$ degrees of freedom. It is clearly visible that the 
	   LR-test (red) deviates from the uniform expectation for small p-values. This means that the LR-test severely overestimates the significance in this case. The distribution of p-values 
	   for the F-test (blue) is compatible with the uniform expectation. The distribution of p-values in the lower panel is generated when the simulated spectral points are normal distributed around 
	   their true value. In this case, the p-value distribution for both tests is compatible with the uniform expectation. The parameters of the Monte Carlo simulation are discussed in Sec. 
	   \ref{robustness_section}.}
	 \label{pvalue_dis}
\end{figure*}

Consider a powerlaw model for the $\gamma$-ray flux $\phi(E)$ at energy $E$ with an exponential cutoff at $E_\mathrm{cut,\;\gamma}=1/\lambda_\gamma$. The model can be parameterized by 
\begin{equation}
	\phi(E) = \phi_0 \left( \frac{E}{E_0} \right)^{-\Gamma}\exp(-(\lambda_\gamma E)^\beta)
	\label{ecpl}
\end{equation}
where again $\Gamma$ is the spectral index and $\phi_0$ the flux normalization at energy $E_0$. The parameter $E_0$ is assumed to be fixed to $1$ TeV in the following. Also fixed is the parameter 
$\beta$ which controls the shape of the exponential cutoff. Frequently discussed is, for example, the case $\beta=1$ (see e.g. \cite{casA}). The hadronic model given by Eq. \ref{hadron_spectrum} 
reduces to Eq. \ref{ecpl} with $\beta=0.5$ and $\lambda_\gamma=16\lambda_h$. The cutoff parameter $\lambda_\gamma$ is constrained to $\lambda_\gamma>0$ because a $\gamma$-ray flux suppression 
for energies above $E_\mathrm{cut,\;\gamma}$ is expected for Pevatrons. 
The model given by Eq. \ref{ecpl} has three free parameters $\boldsymbol{\theta_1}=(\phi_0,\;\Gamma,\;\lambda_\gamma)$.
The question is whether the powerlaw model with exponential cutoff (Eq. \ref{ecpl}) gives a better description of a dataset than a powerlaw model parameterized by 
\begin{equation}
	\phi_\mathrm{PL}(E) = \phi_0 \left( \frac{E}{E_0}\right)^{-\Gamma}\;\mathrm{.}
	\label{pl}
\end{equation}
This model has two free parameters $\boldsymbol{\theta_0}=(\phi_0,\;\Gamma)$ while $E_0$ is again fixed to $1$ TeV. 
To select the best fitting model, the null hypothesis $\mathrm{H_0}$, i.e. the absence of an exponential cutoff ($\lambda_\gamma=0$),
can be tested against the physically constrained alternative hypothesis $\mathrm{H_1}$, $\lambda_\gamma>0$, by means of a hypothesis test at a given confidence level $\mathrm{CL}$.\\
Let in the following $\phi_i$, $i=0,\dots, N-1$, denote binned $\gamma$-ray flux measurements with bin-centers at energies $E_i$. All $N$ measurements of $\phi_i$ are assumed to be independent. 
In practice, the independence of the $\phi_i$ 
can be expected when the energy bin-width for the calculation of $\phi_i$ is much larger than the instrumental energy resolution. Let further $\sigma_i$ be the measurement errors 
corresponding to the $\gamma$-ray fluxes $\phi_i$. The measurements $\phi_i$ are, in a first step, assumed to be normal distributed around a hypothetical true value $\hat{\phi}_i$. 
A model $\phi(E_i|\boldsymbol{\theta})$ with parameters $\boldsymbol{\theta}$ predicts the true flux $\hat{\phi}_i$. 
Two frequently used tests for a powerlaw hypothesis (Eq. \ref{pl}) against a model with exponential cutoff (Eq. \ref{ecpl}) are presented in the following. 
Afterwards, the robustness of the tests against deviations from the assumed normal error model is discussed.
\subsubsection{Likelihood ratio test}
\label{lrtest}
Because the flux measurements are assumed to be independent, the likelihood function for the parameters $\boldsymbol{\theta}$ of a given model factorizes to 
$L(\boldsymbol{\theta})=\prod_{i=0}^{N-1}\frac{1}{\sqrt{2\pi\sigma_i^2}}\exp\left(-\frac{(\phi_i-\phi(E_i|\boldsymbol{\theta}))^2}{2\sigma_i^2}\right)$. 
Let $\boldsymbol{\hat{\theta}_k}=\mathrm{argmax}_{H_k}L(\boldsymbol{\theta})$ be the maximum likelihood estimates for the model parameters under the hypotheses $\mathrm{H_k}$ with $k\in\{0,1\}$. 
Consider the LR-test with the statistic 
$-2\ln\Lambda=-2\ln(L(\boldsymbol{\hat{\theta}_0})/L(\boldsymbol{\hat{\theta}_1}))=\mathrm{RSS}(\boldsymbol{\hat{\theta}_0})-\mathrm{RSS}(\boldsymbol{\hat{\theta}_1})$ where
\begin{equation}
	\label{rss}
	\mathrm{RSS}(\boldsymbol{\hat{\theta}_k})=\sum_{i=0}^{N-1}\left(\frac{\phi_i-\phi(E_i|\boldsymbol{\hat{\theta}_k})}{\sigma_i}\right)^2 
\end{equation}
for the hypotheses $\mathrm{H_k}$. The null hypothesis is discarded at confidence level $\mathrm{CL}$ 
if $-2\ln\Lambda>\Lambda_\mathrm{crit}(\mathrm{CL})$. The critical value $\Lambda_\mathrm{crit}$ can be calculated using the results from \cite{chernoff} 
because the two models given by Eqs. \ref{pl} and 
\ref{ecpl} are nested, i.e. Eq. \ref{ecpl} reduces to Eq. \ref{pl} when $\lambda_\gamma\to 0$. 
When $F_\mathrm{\chi^2_1}^{-1}$ denotes the inverse cumulative density function of a $\chi^2$-distributed random variable with one degree of freedom, the critical value is given by 
$\Lambda_\mathrm{crit}=F_\mathrm{\chi^2_1}^{-1}(2\mathrm{CL}-1)$, see appendix \ref{critical_value_appendix}. The test will have the expected false positive error rate 
$1-\mathrm{CL}$ when the normal error model holds and the sample size $N$ is large such that $\phi(E_i|\boldsymbol{\hat{\theta}_k})\to\hat{\phi}_i$ when $\mathrm{H_0}$ is true.
\subsubsection{F-test}
\label{ftest}
The F-test is, in the case of $N$ independent measurements and two nested models with $2$ and $3$ parameters, based on the test statistic
\begin{equation}
	F=(N-3)\frac{\mathrm{RSS}(\boldsymbol{\hat{\theta}_0})-\mathrm{RSS}(\boldsymbol{\hat{\theta}_1})}{\mathrm{RSS}(\boldsymbol{\hat{\theta}_1})}\;\mathrm{.}
	\label{fe0}
\end{equation}
The null hypothesis $\lambda_\gamma=0$ is discarded in favor of the alternative hypothesis $\lambda_\gamma>0$ 
at confidence level CL when $F>F_\mathrm{crit}(\mathrm{CL})$. Similar to the LR-test, the critical value is given by 
\begin{equation}
	\label{f_crit}
	F_\mathrm{crit}=F_{1,N-3}^{-1}(2\mathrm{CL}-1)
\end{equation}
where $F^{-1}_{1,N-3}$ is the inverse cumulative density function of a F-distributed random variable with $1$ (numerator) and $N-3$ (denominator) degrees of freedom. 
The F-test is exact for small samples $N$ when linear models are being compared. For non-linear models, 
such as in the considered case, the false positive error rate can only be expected to be asymptotically \citep{nonlinear_article}. 
Like the special LR-test constructed above, the F-test relies on the assumption of a normal error model.
\subsection{Robustness of the LR- and F-tests}
\label{robustness_section}
The upper plot in Fig. \ref{best_and_worst_spectrum} shows spectral data points from the HGPS for the $\gamma$-ray source HESS J1800-240. A fit of the data to a powerlaw model 
(Eq. \ref{pl}) results in the best fit parameters $\boldsymbol{\theta_\mathrm{s}}=(\phi_0,\;\Gamma)=((4.3\pm 0.1)\cdot 10^{-13}\;\mathrm{TeV^{-1}cm^{-2}s^{-1}}, 2.44\pm0.02)$. 
The robustness of the LR- and F-test constructed above is tested in a Monte Carlo simulation of true null models 
where random spectra are created based on the best fit model $\phi(E|\boldsymbol{\theta_\mathrm{s}})$. 
For each simulated spectrum, all five spectral points are sampled from a Student t-distribution $t_5(\mathrm{location},\;\mathrm{scale})$ 
with location parameter $\phi(E_i|\boldsymbol{\theta_\mathrm{s}})$, scale parameter $\sigma_i$ and $5$ degrees of freedom. 
For comparison, the same simulation is repeated 
by sampling from a Normal distribution with mean $\phi(E_i|\boldsymbol{\theta_\mathrm{s}})$ and standard deviation $\sigma_i$. 
The Student error model predicts more flux outliers than expected in the 
normal error model on which the constructed LR- and F-tests rely.\\
Figure \ref{pvalue_dis} shows the distribution of LR- and F-test p-values for Monte Carlo simulations with the Student (upper panel) and normal (lower panel) error model. 
The shown p-values refer to tests of a powerlaw model against a model with exponential cutoff in the $\gamma$-ray spectrum (Eq. \ref{ecpl} with $\beta=1$). Similar simulations with unchanged 
conclusions were performed when a powerlaw model is tested against a model where the $\gamma$-ray emission is modeled in a hadronic scenario (Eq. \ref{hadron_spectrum}). 
The lower panel of Fig. \ref{pvalue_dis} shows that the distribution of p-values for a true null hypothesis is compatible with the uniform expectation when the normal error model is used. However, 
severe discrepancies between the uniform expectation and the LR-test are observed in the upper panel of Fig. \ref{pvalue_dis}. In contrast, 
the F-test shows no indication for a deviation from the uniform expectation when the normal error model is not true. The bias towards low p-values of the LR-test 
in the Student error model is equivalent to a bias towards larger significances. Similar Monte Carlo simulations for all other $\gamma$-ray sources that were selected for the analysis (see 
Sec. \ref{all_data}) confirm the observation that the LR-test is not robust against violations of the normal error model in the HGPS catalog.\\
The improved robustness of the F-test compared to the LR-test 
comes at the price of a reduced test power. The test power can be estimated with a toy Monte Carlo simulation of true alternative hypotheses. 
For this, five spectral points are calculated at energies $E_i=\xi^iE_\mathrm{min}$ where 
$\xi=2.4$ and $E_\mathrm{min}=0.4$ TeV are the median values of the binning factor and the minimal energy bin
for the selected HGPS data (see Tab. \ref{sel_hgps_data}). At each energy, the flux is sampled from 
$\phi_0(E/E_0)^{-\Gamma}\exp(-E/E_\mathrm{cut,\;\gamma})$ where $\phi_0=3.51\cdot 10^{-12}\;\mathrm{cm^{-2}s^{-1}TeV^{-1}}$ and $\Gamma=2.11$ are the median flux normalization and 
powerlaw index of the selected HGPS data (see Tab. \ref{sel_hgps_data}). The flux normalization energy $E_0=1$ TeV is fixed and the cutoff energy $E_\mathrm{cut,\;\gamma}$ is varied. 
The error of each sampled flux point is set to $20\%$ of the flux in this toy simulation. Figure \ref{coverage_dis} compares 
the energy cutoff detection power between the LR- and the F-test. It is clearly seen that there is no true energy cutoff 
where the F-test has more power than the LR-test. However, for example for a true energy cutoff of $100$ TeV, the LR-test has $10\%$ more 
power than the F-test.\\
Despite the reduced power of the F-test when compared to the LR-test, the possible influence of systematic effects in HGPS data motivates the usage of the F-test in the following discussion.
\subsection{Lower limit on the energy cutoff}
\label{confidence_interval_section}
\begin{figure}
	\centering
	\includegraphics[width=\hsize]{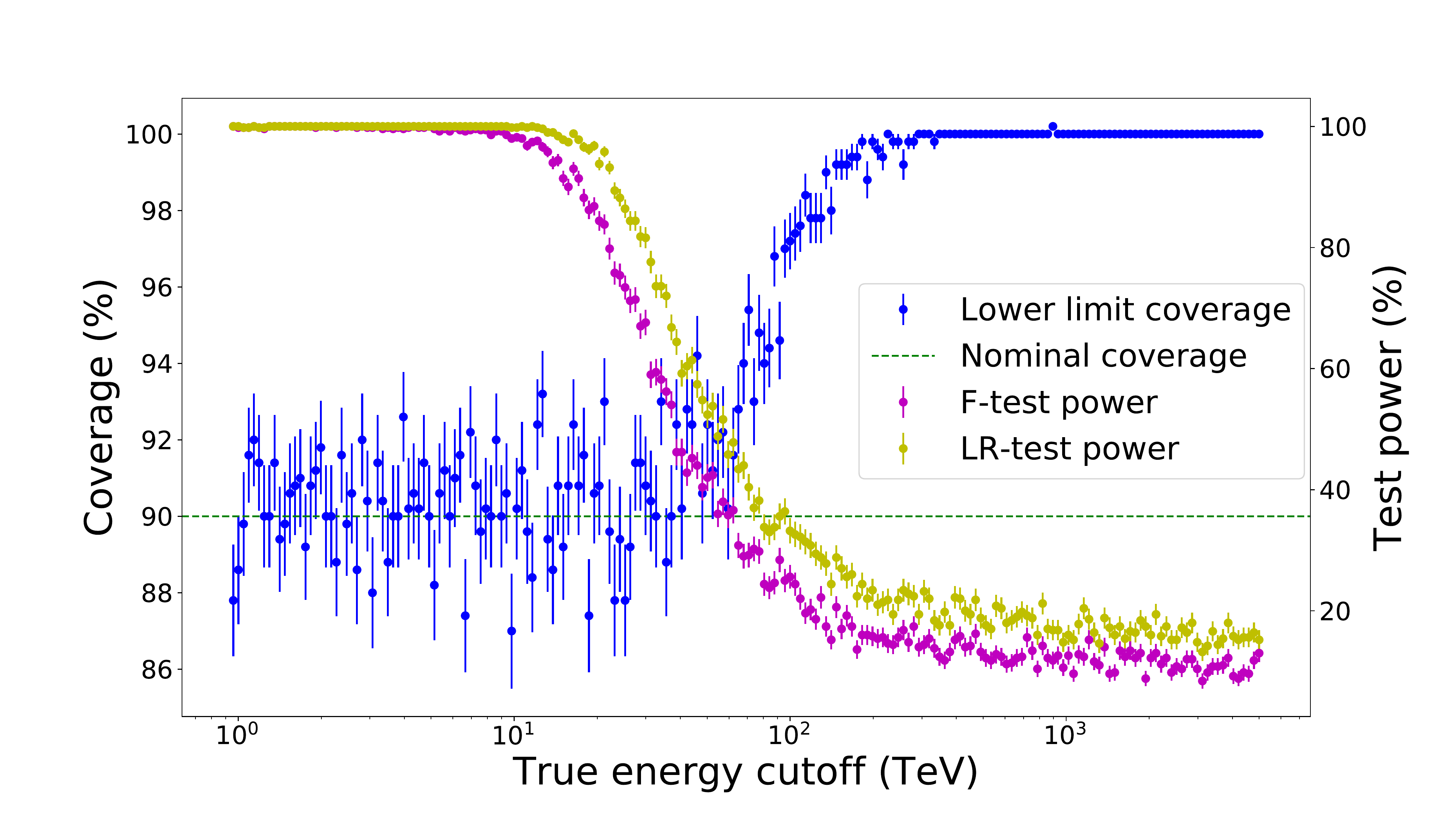}
	\caption{Shown in blue is the simulated coverage of the interval $I$ given by Eq. \ref{lower_limit} at $90\%$ CL. 
	The coverage is compatible with the nominal confidence level for true energy cutoffs below $50$ TeV where the F-test power (magenta) is large. The interval $I$ overcovers for large true energy 
	cutoffs where the F-test power is small. The LR-test power is, for comparison, shown in yellow.}
	\label{coverage_dis}
\end{figure}

In practice a hypothesis test for the presence of a cutoff is performed at high confidence level, e.g. at a significance level of $5\sigma$. If the powerlaw 
hypothesis is not discarded, a lower limit on the energy cutoff $E_\mathrm{cut}$ is requested at reduced confidence level (e.g. $90\%$) 
to constrain theoretical models. Depending on the alternative hypothesis, $E_\mathrm{cut}$ can refer to $E_\mathrm{cut,\;\gamma}$ (Eq. \ref{ecpl}) or to $E_\mathrm{cut,\;h}$ (Eq. \ref{hadron_spectrum}). 
A lower limit on the energy cutoff corresponds to a one-sided confidence interval. Let $E_\mathrm{cut}^\mathrm{MLE}$ be the maximum likelihood estimate 
of the energy cutoff as determined in a least-square fit. Consider the interval 
\begin{equation}
	I=(E_\mathrm{cut}^{-},\infty)
	\label{lower_limit}
\end{equation}
where $E_\mathrm{cut}^{-}$ is the solution to the equation 
\begin{equation}
	F(E)=F_\mathrm{crit}
	\label{limit_eq}
\end{equation}
in the interval $0<E<E_\mathrm{cut}^{\mathrm{MLE}}$. 
Here, 
\begin{equation}
	\label{fe}
	F(E)=(N-3)\frac{\mathrm{RSS}(\boldsymbol{\hat{\theta}_0^{*}}|E)-\mathrm{RSS}(\boldsymbol{\hat{\theta}_1})}{\mathrm{RSS}(\boldsymbol{\hat{\theta}_1})}
\end{equation}
and $\mathrm{RSS}$ is given by Eq. \ref{rss}. The parameters $\boldsymbol{\hat{\theta}_0^{*}}$ are the maximum likelihood estimators of $\phi_0$ and $\Gamma$ 
for a powerlaw model with fixed energy cutoff at $E$. 
Similarly, $\boldsymbol{\hat{\theta}_1}$ is the maximum likelihood estimator for the parameters of a powerlaw model with variable exponential energy cutoff, 
again with the constraint that $E_\mathrm{cut}>0$. The critical value $F_\mathrm{crit}=F_\mathrm{crit}(\mathrm{CL})$ is given by Eq. \ref{f_crit}.\\
Equation \ref{fe} is the result of an F-test of the hypothesis $\mathrm{H_0}:\; E_\mathrm{cut}=E$ against the 
alternative hypothesis $\mathrm{H_1}:\; E_\mathrm{cut}\neq E\;\mathrm{and}\; E_\mathrm{cut}>0$. The existence of $E_\mathrm{cut}^{-}$, defined in Eq. \ref{lower_limit}, is shown in 
appendix \ref{existence}. In the following, it is assumed that $E_\mathrm{cut}^{-}$ is the unique solution of Eq. \ref{fe} in the energy interval $0<E<E_\mathrm{cut}^\mathrm{MLE}$ 
(see also the discussion in appendix \ref{existence}).\\
It is shown in appendix \ref{lower_limit_appendix} that the interval $I$ 
is expected to overcover when the true energy cutoff is large. However, excellent frequentist coverage properties are expected when the true energy cutoff is small such that the F-test power 
at the given CL is large. Figure \ref{coverage_dis} shows the result of a Monte Carlo simulation where $\gamma$-ray energy spectra with energy cutoff shape $\beta=1$ were 
simulated following the respective discussion in Sec. \ref{robustness_section}. The result of this simulation shows that the frequentist coverage of the interval $I$ agrees with the nominal 
$90\%$ confidence level 
when the true energy cutoff is smaller than $50$ TeV. For energy cutoffs larger than $50$ TeV, the test power at the 
confidence level of $90\%$ decreases fast and the interval overcovers.\\
It is concluded that the lower limit $E_\mathrm{cut}^{-}$ defined in Eq. \ref{lower_limit} has very good coverage properties when the true cutoff of the $\gamma$-ray spectrum 
is small. Otherwise it is conservative. 
For typical source and instrumental parameters representative for the HESS galactic plane survey, good coverage can be expected when the true cutoff of the $\gamma$-ray energy spectrum is below 
$50$ TeV.

\subsection{Systematic errors}
The lower limit on the energy cutoff is the lower end of the confidence interval $I$, given by Eq. \ref{lower_limit}. This is a confidence interval for 
the maximum likelihood estimator $E_\mathrm{cut}^\mathrm{MLE}$ of the energy cutoff. Systematic effects on the 
energy cutoff become dominant when the systematic variation of $E_\mathrm{cut}^\mathrm{MLE}$ is not covered by the confidence interval $I$. For data from Cherenkov telescope arrays like HESS and 
VERITAS, reasons for systematic variations of $E_\mathrm{cut}^\mathrm{MLE}$ can be, for example, atmospheric and analysis effects or instrumental problems like broken camera elements. 
In case of the HGPS, 
the systematic errors on the fit parameters of a powerlaw model are estimated to be $30\%$ on the flux normalization and $0.2$ on the powerlaw index \citep{hess_plane}. 
These results were obtained from the comparison of the fit results obtained with different analysis methods. 
However, no public information is available on the systematic variation of $E_\mathrm{cut}^\mathrm{MLE}$ in the HGPS. 
\section{Analysis of HGPS data}
\label{hess_survey_section}
The analysis of the HGPS data, after the quality selection discussed in Sec. \ref{data_selection}, starts with a consistency check. It is ensured that the HGPS spectral points can be 
fit with results that are in reasonable agreement with the HGPS analysis.
Afterwards, lower limits on the spectral cutoffs are derived for the selected $\gamma$-ray sources.\\
The HGPS catalog contains best fitting powerlaw parameters for each source. The fit in the HGPS analysis is performed with an unbinned forward folding method \citep{forward_folding}. In contrast, 
binned least square fits of the public spectral HGPS data are used in this analysis because unbinned data are not publicly available. The best least square fit parameters are, 
however, in reasonable agreement with the cataloged best fit parameters. 
The maximum absolute difference between the powerlaw index obtained in this analysis and the powerlaw 
index given the HGPS catalog is $1.8\sigma$ for HESS J1708-443 where $\sigma$ denotes the HGPS catalog index error. 
The median absolute difference between the fitted powerlaw indices is $0.6\sigma$. 
The corresponding largest absolute difference between the flux normalizations is $1.2\sigma$ for HESS J1026-582. The median absolute flux difference for the dataset is $0.1\sigma$.\\
Table \ref{analysis_result} shows the results of a search for spectral cutoffs for the $25$ selected $\gamma$-ray sources in the HGPS. The best fit exponential cutoff models (Eq. \ref{ecpl} with 
$\beta=1$ and Eq. \ref{hadron_spectrum}) have a p-value larger than $1\%$ in a $\chi^2$ test for all considered spectra. All calculated solutions to Eq. \ref{fe} are unique. 
No energy cutoff is detected with a statistical 
significance exceeding $5\sigma$ for any of the analyzed sources. The lower limit on the energy cutoff in a hadronic scenario is larger than $100$ TeV for five sources. For these five sources, 
the best fit powerlaw index is in the range between $2.0$ and $2.5$. Figure \ref{best_and_worst_spectrum} shows the spectral data together with the best fit powerlaw and the excluded cutoffs for the 
sources HESS J1800-240 and HESS J1746-285. For these sources, the most and least constraining lower limit on the energy cutoff is derived in this analysis.\\
No systematic error on the lower limit on the energy cutoff can be derived based on HGPS data alone since the systematic variation of $E_\mathrm{cut}^\mathrm{MLE}$ is not public. However, 
in case of the source HESS J1908+063, an at least partially independent dataset is available. This dataset is discussed in the next section.

\onecolumn
\begin{longtable}{lrrrrrrrr}
	\caption{Results of the energy cutoff analysis for $25$ unidentified $\gamma$-ray sources in the HGPS catalog: The source name refers to the HGPS \citep{hess_plane}. 
	$E_\mathrm{cut,\;\gamma}$ is the lower limit on the exponential energy cutoff of 
	the $\gamma$-ray spectrum (Eq. \ref{ecpl} with $\beta=1$). $\mathrm{S_\gamma}$ is the significance of an F-test for a powerlaw model against 
	an exponential cutoff in the $\gamma$-ray spectrum (again modeled by Eq. \ref{ecpl} with $\beta=1$). $E_\mathrm{cut,\;h}$ is the upper limit on the energy cutoff of a 
	hadron population in a purely hadronic model (Eq. \ref{hadron_spectrum}). 
	$E_\mathrm{max}$ is the maximum bin-center energy of the spectral points for the source used in the analysis. $\Gamma$ is the powerlaw index of the best least square 
	fit to the public HGPS data. Errors are statistical only. 
	N is the number of spectral points for the source used in the analysis. The p-values refers to a $\chi^2$ goodness of fit test for the powerlaw model. 
	The HESS livetime used for the spectral analysis of the source in the HGPS is indicated by T. 
	All upper limits are at $90\%$ CL. The results are ordered descending in $E_\mathrm{cut,\;h}$.}\\
	\hline\hline

	\label{analysis_result}
	Source &  $E_\mathrm{cut,\;\gamma}$ (TeV) &  $\mathrm{S_\gamma}$ &  
	$E_\mathrm{cut,\; h}$ (TeV) &  $E_\mathrm{max}$ (TeV) & $\Gamma$ &  N &  p-value & T (h)\\
	\hline
	HESS J1800-240 &                             34.7 &    0.0 &                             510.1 &           12 &  2.44$\pm$0.03 &  5 &   9.6e-01 &     10 \\
	HESS J1641-463 &                             21.3 &    0.1 &                             184.9 &           13 &  2.38$\pm$0.03 &  5 &   9.8e-01 &     27 \\
	HESS J1908+063 &                             31.1 &    1.0 &                             154.0 &           30 &  2.22$\pm$0.04 &  6 &   8.2e-01 &     12 \\
	HESS J1852-000 &                             29.2 &    0.0 &                             119.7 &           30 &  2.12$\pm$0.06 &  6 &   8.0e-01 &     23 \\
	HESS J1634-472 &                             20.3 &    0.5 &                             107.8 &           30 &  2.30$\pm$0.05 &  6 &   2.8e-01 &     14 \\
	HESS J1828-099 &                             14.5 &    0.0 &                              64.6 &           12 &  2.19$\pm$0.07 &  5 &   8.3e-01 &     20 \\
	HESS J1023-575 &                             13.5 &    2.0 &                              35.3 &           32 &  2.38$\pm$0.07 &  6 &   2.1e-01 &     23 \\
	HESS J1858+020 &                             10.6 &    0.3 &                              33.5 &           13 &  2.30$\pm$0.04 &  5 &   9.5e-01 &     22 \\
	HESS J1841-055 &                              8.4 &    2.7 &                              29.9 &           12 &  2.42$\pm$0.07 &  5 &   9.8e-03 &     14 \\
	HESS J1503-582 &                              8.3 &    0.0 &                              28.4 &           14 &  2.7$\pm$0.1 &  5 &   3.4e-01 &     18 \\
	HESS J1507-622 &                             10.6 &    2.4 &                              26.3 &           15 &  2.10$\pm$0.07 &  5 &   4.5e-01 &     13 \\
	HESS J1646-458 &                              7.4 &    0.0 &                              22.7 &           13 &  2.6$\pm$0.1 &  5 &   7.1e-01 &     10 \\
	HESS J1457-593 &                              7.7 &    0.0 &                              18.1 &           15 &  2.6$\pm$0.1 &  5 &   7.3e-01 &      4 \\
	HESS J1843-033 &                              9.9 &    2.8 &                              16.6 &           30 &  2.2$\pm$0.1 &  6 &   1.7e-03 &     17 \\
	HESS J1018-589 B &                              8.6 &    2.3 &                              14.7 &           15 &   2.1$\pm$0.1 &  5 &   6.3e-01 &     21 \\
	HESS J1632-478 &                              5.3 &    0.1 &                              10.0 &           13 &  2.46$\pm$0.06 &  5 &   4.0e-01 &     15 \\
	HESS J1741-302 &                              4.5 &    0.0 &                               9.2 &            7 &  2.2$\pm$0.2 &  6 &   5.3e-01 &    145 \\
	HESS J1702-420 &                              5.2 &    0.2 &                               9.1 &           13 &   2.1$\pm$0.1 &  5 &   1.7e-01 &      4 \\
	HESS J1809-193 &                              3.4 &    1.0 &                               5.0 &           12 &   2.3$\pm$0.1 &  5 &   1.6e-01 &     10 \\
	HESS J1616-508 &                              3.2 &    1.1 &                               3.8 &           13 &  2.3$\pm$0.1 &  5 &   1.1e-02 &      4 \\
	HESS J1808-204 &                              2.1 &    0.0 &                               2.7 &           12 &  2.1$\pm$0.2 &  5 &   1.3e-01 &     37 \\
	HESS J1026-582 &                              5.0 &    2.0 &                               2.3 &           32 &  2.0$\pm$0.2 &  5 &   7.0e-03 &     22 \\
	HESS J1804-216 &                              1.3 &    1.1 &                               1.8 &           12 &  2.7$\pm$0.1 &  5 &   2.6e-04 &     19 \\
	HESS J1708-443 &                              2.2 &    1.2 &                               1.3 &           13 &  2.0$\pm$0.2 &  5 &   4.2e-02 &      8 \\
	HESS J1746-285 &                              1.2 &    0.6 &                               0.7 &            6 &  2.0$\pm$0.2 &  7 &   4.0e-01 &    259 \\

	\hline
\end{longtable}
\twocolumn

\begin{figure*}
	\centering
	\includegraphics[width=\hsize]{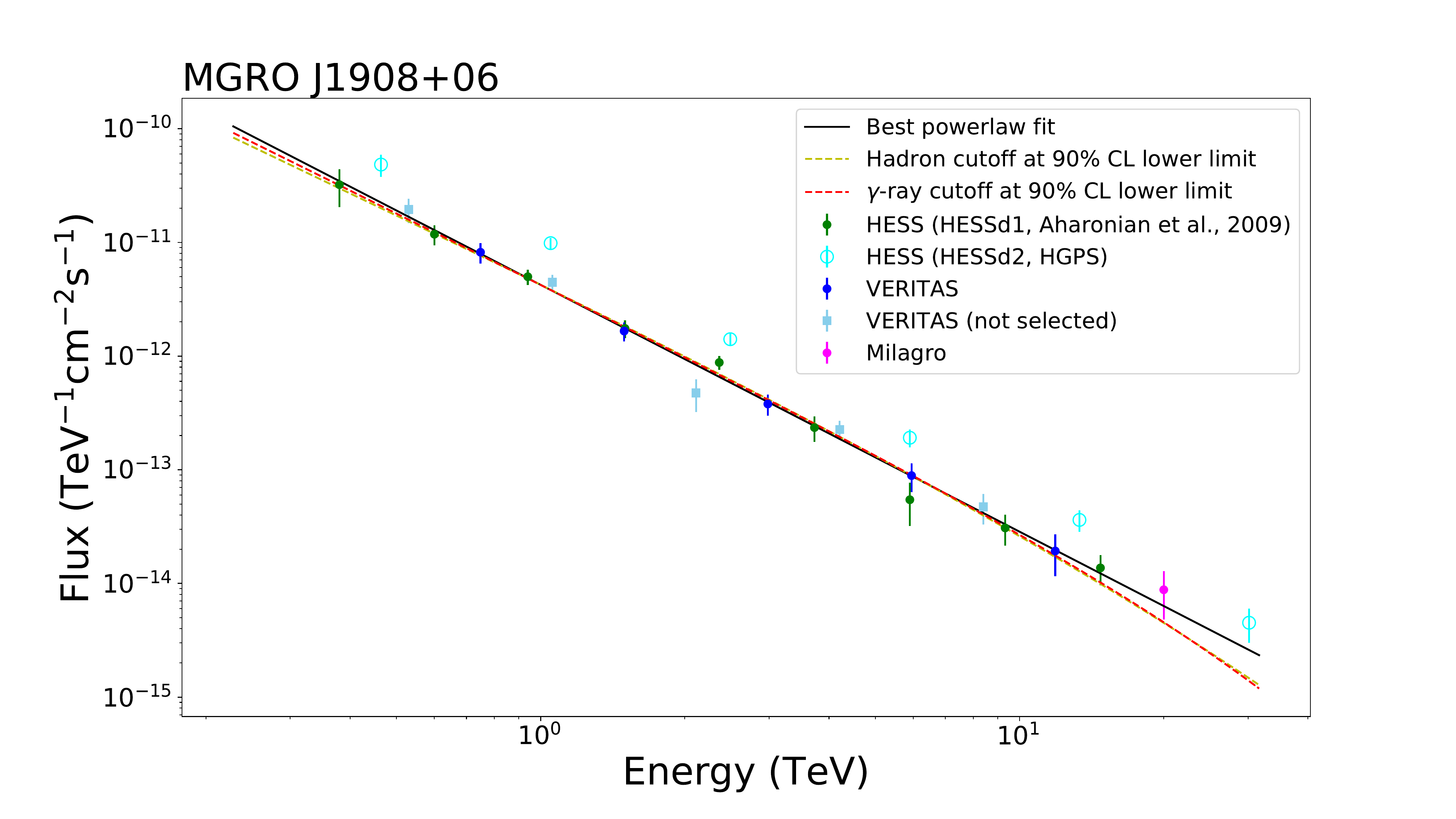}
	\caption{Spectral $\gamma$-ray data for the source MGRO J1908+06. Shown as green circles is the HESS data from \cite{hess1908}. Shown in cyan open circles is the spectral data from the HGPS. 
	Blue spectral points are from VERITAS \citep{veritas1908}. VERITAS data marked with a square is not selected for the analysis to avoid correlations between spectral points. 
	The Milagro data point is from \cite{milagro1908}. Shown in black, yellow and red are best fits to all datapoints indicated 
	by filled circles. The black line is a fit to a powerlaw model. Yellow and red lines are best fits to a powerlaw model with cutoff at the $90\%$ CL 
	lower limits $E_\mathrm{cut,\;\gamma}=29.5$ TeV (yellow) and $E_\mathrm{cut,\;h}=141.5$ TeV (red).}
	\label{j1908}
\end{figure*}

\section{Analysis of MGRO J1908+06 data}
\label{j1908_section}

The $\gamma$-ray source MGRO J1908+06 is associated with HESS J1908+063. This $\gamma$-ray source is one of the unidentified HGPS sources for which the lower limit on the energy cutoff in a 
hadronic scenario is found to be larger than $100$ TeV in section \ref{hess_survey_section}. Compared to the HGPS catalog, additional public 
data for this $\gamma$-ray source is available. This data, acquired with different experiments, is described and analyzed in this section.
\subsection{Data for MGRO J1908+06}
Milagro detected the $\gamma$-ray source at a median energy of $20$ TeV with a flux of $(8.8\pm 2.4)\cdot 10^{-15}\;\mathrm{TeV^{-1}cm^{-2}s^{-1}}$ \citep{milagro1908}. 
Based on an analysis of $27$h data acquired with the HESS experiment, spectral data is published in \cite{hess1908}. This dataset will be denoted by HESSd1 in the following while HESSd2 will be 
used to identify the spectral HGPS data on MGRO J1908+06. The $27$h HESSd1 dataset must be at least partially independent from the $12$h HESSd2 because the livetime is larger. 
VERITAS also observed MGRO J1908+063 and collected $62$h 
data between the years $2007$ and $2012$. Spectral datapoints derived from the VERITAS dataset are published in \cite{veritas1908}. The spectral properties of the HESSd1, 
VERITAS and Milagro dataset are compatible with each other \citep{veritas1908}. Two lower limits on the energy cutoff $E_\mathrm{cut,\;\gamma}$ of the $\gamma$-ray spectrum 
were derived before. A $90\%$ CL 
lower limit on $E_\mathrm{cut,\;\gamma}$ is derived from a combination of the VERITAS and Milagro data at $17.7$ TeV \citep{veritas1908}. 
Also at $90\%$ CL, the combination of the HESSd1 and Milagro data leads to a lower limit of $E_\mathrm{cut,\;\gamma}=19.1$ TeV \citep{hess1908}.\\
The additional Cherenkov telescope data for MGRO J1908+063 from HESS and VERITAS must pass the same quality selection as discussed for the HGPS spectral data in section \ref{data_selection}. 
The HESSd1 dataset passes this selection. However, the public spectral data for MGRO J1908+063 from VERITAS is binned with a factor $\xi=1.4$ while the energy resolution is 
similar to HESS \citep{veritas}. Only every second spectral point from the public VERITAS data is therefore used in this analysis.
The following analysis results are checked with the full VERITAS dataset to ensure that no bias towards better results is introduced with this spectral point selection.

\subsection{Data analysis}
\begin{figure}
	\centering
	\includegraphics[width=\hsize]{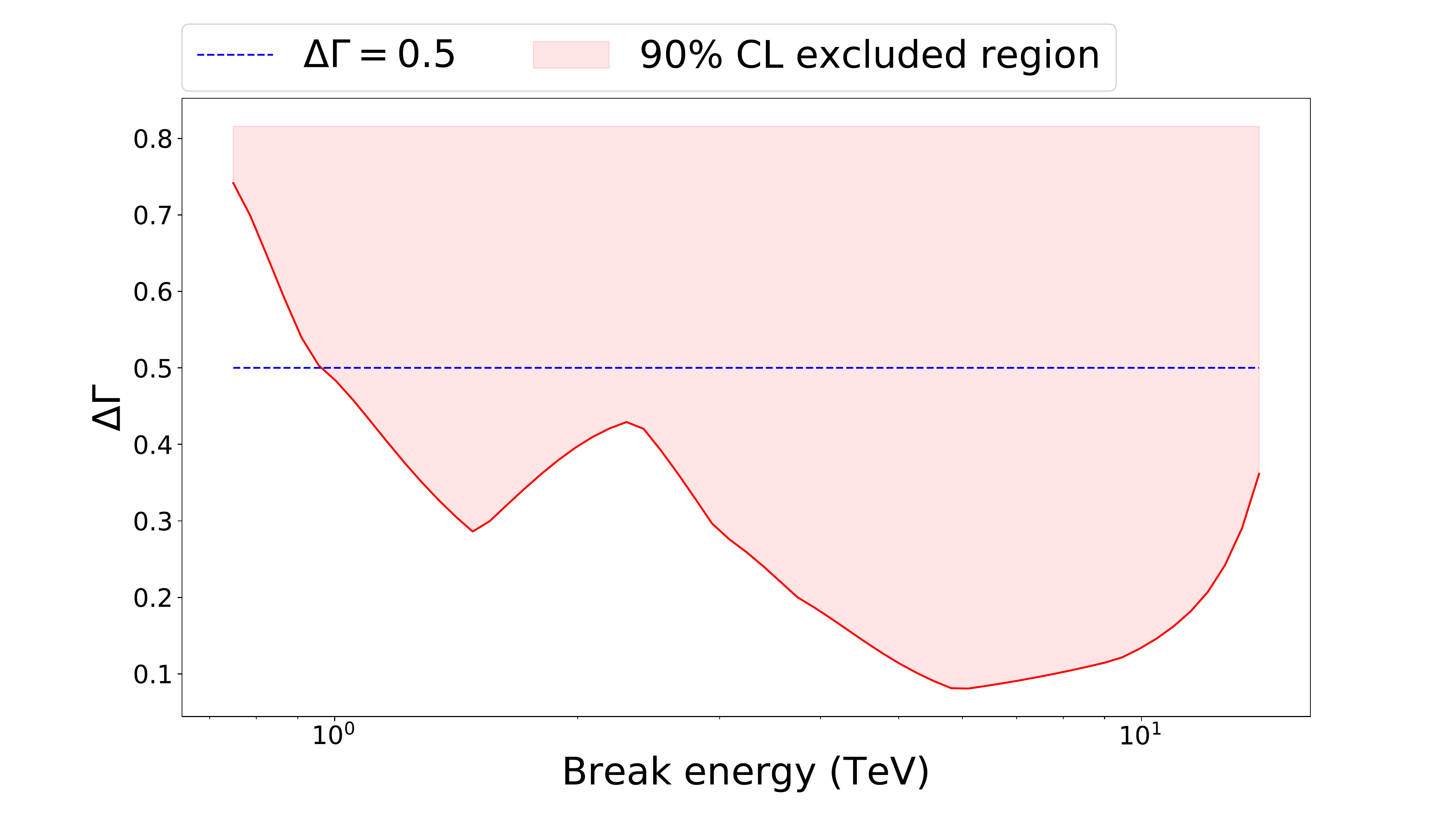}
	\caption{Upper limit on a break in the powerlaw spectrum by $\Delta\Gamma$ as a function of the break energy. Shown as red line is the $90\%$ CL upper limit on $\Delta\Gamma$ derived 
	from the combined data discussed in Sec. \ref{j1908_section}.}
	\label{j1908_break}
\end{figure}

The analysis of the selected spectral data from HESSd1, VERITAS and Milagro results in a best fit powerlaw spectrum with 
$\phi_0=(4.2\pm0.2)\cdot 10^{-12}\;\mathrm{cm^{-2}s^{-1}TeV^{-1}}$ and $\Gamma=2.17\pm0.04$. The powerlaw index is compatible with the index $\Gamma=2.26\pm0.06$ cataloged in the HGPS while the 
flux normalization in the HGPS, $(1.05\pm 0.09)\cdot 10^{-11}\;\mathrm{cm^{-2}s^{-1}TeV^{-1}}$, is incompatible. The 
fit to a powerlaw model has a p-value of $0.91$ in a $\chi^2$ goodness of fit test. 
The lower $90\%$ CL limits on the spectral energy cutoffs derived from this dataset are $E_\mathrm{cut,\;\gamma}=29.5$ TeV and $E_\mathrm{cut,\;h}=141.5$ TeV. 
These limits confirm the limits derived from the HGPS data. Overall, the agreement between this analysis and the HGPS analysis result is very good in regard to the derived lower limits on 
spectral energy cutoffs. Data and best fitting models are shown in Fig. \ref{j1908}.\\
Compared to the HESSd2 dataset with $6$ spectral points, the combined data from HESSd1, VERITAS and Milagro considered in this section contains with $15$ spectral points many more 
flux measurements. This allows to search for a spectral break in the TeV energy 
range. Multiple F-tests are performed where the fit to the data of a powerlaw model ($H_0:\;\Delta\Gamma=0$) and a broken powerlaw ($H_1:\;\Delta\Gamma>0$, see Eq. \ref{broken_pl}) are compared. 
The comparison cannot be performed in a single test because the break energy $E_\mathrm{break}$ is not defined under the null hypothesis. 
Let $E_i$ again be the energies of the $N$ available spectral flux data points. The energy range $E_2=0.75$ TeV to $E_{N-2}=14.8$ TeV is scanned for a break energy $E_\mathrm{break}$ 
with a logarithmic binning factor of $1.05$. The F-test for the comparison of the fit quality of the powerlaw model and the broken powerlaw model, assuming the respective 
energy break energy, is performed in each scanning step. The F-test is inverted to derive an upper limit on the index change $\Delta\Gamma$ as a function of the 
assumed break energy $E_\mathrm{break}$. The result is shown in Fig. \ref{j1908_break}. An index change $\Delta\Gamma>0.5$ is ruled out at $90\%$ CL for energies between $1$ TeV and $10$ TeV.\\
The largest local significance obtained in the $61$ tests of the energy break scan is $p_\mathrm{local}=0.14$ with a F-test statistic $c=1.23$. 
The local significance must be transformed into a global significance $p_\mathrm{global}$ considering the number of 
performed tests. Based on results for the global p-value in a multiple testing scenario (see \cite{gross_vitells} and references therein), \cite{algeri_f} derive 
a global p-value correction for an F-process. This correction reads in the case considered here
\begin{equation}
	p_\mathrm{global} \leq p_\mathrm{local} + \left(\frac{N-3+c}{N-3+c_0}\right)^{-N/2+2}E[N_{c_0}]\;\mathrm{.}
	\label{global}
\end{equation}
$E[N_{c_0}]$ is the expected number of upcrossings over the level $c_0$ of the process of F-test statistics during the energy cutoff scan when the null hypothesis is true. Equation \ref{global} 
allows to extrapolate the global p-value correction from a low test statistic level $c_0$, 
where a small number of simulations is required for the estimation of $E[N_{c_0}]$, to a larger value $c$ of the test statistic. 
Following \cite{multiple_jan}, $E[N_{c_0}]$ is estimated in a parametric bootstrap simulation at $c_0=0.3$ to be $E[N_{c_0}]=0.52\pm0.02$. With this correction, the global p-value 
of the F-test for a spectral break in the selected MGRO J1908+06 data from HESS, VERITAS and Milagro can be estimated to be $p_\mathrm{global}=0.5$. The powerlaw hypothesis is therefore not 
discarded in favor of a broken powerlaw.

\section{Discussion}
\label{discussion_section}
The analysis of HGPS data resulted in five $\gamma$-ray sources for which the energy spectrum is compatible with a powerlaw and, in a hadronic scenario, 
the lower limit on the energy cutoff of the accelerated particle population 
is larger than $100$ TeV. These five $\gamma$-ray sources are discussed in detail in this section. An emphasis of the discussion is on the plausibility of the hadronic scenario for these sources. 

\subsection{HESS J1800-240}
The region around HESS J1800-240 is found to have a complex $\gamma$-ray morphology in \cite{j1800}, based on a HESS dataset of $42$h livetime. Four $\gamma$-ray sources are detected in this region. 
The SNR W28 is associated with the source HESS J1801-233. The region around HESS J1800-240 itself is subdivided into three hotspots, spatially coincident with molecular clouds.\\
Based on a smaller dataset of $10$h, the spectral HGPS analysis detects only one $\gamma$-ray source in the region. The lower limit on the energy cutoff for accelerated particles 
derived above is $E_\mathrm{cut,\;h}=510$ TeV at $90\%$ CL. This constraint is comparable to the lower limit of $E_\mathrm{cut,\;h}=600$ TeV at $90\%$ CL 
on the energy cutoff of particles in the vicinity of the Galactic centre in a hadronic scenario \citep{gc}.\\
A physics case for the association of this $\gamma$-ray source with a Pevatron is the model discussed in \cite{snr_mc} where delayed CRs from a SNR illuminate a molecular cloud.
The plausible physics case and the interesting constraint on the energy cutoff in a hadronic scenario may motivate a systematic investigation of differences between the HGPS analysis 
and the analysis in \cite{j1800} as well as a deeper exposure of the region with current Cherenkov telescopes.
\subsection{HESS J1641-463}
The $\gamma$-ray source is discussed in \cite{igor} where the TeV spectrum, extracted from $72$h HESS data, is found to be compatible with a powerlaw model. Additionally, a 
lower limit on the energy cutoff at $E_\mathrm{cut,\;h}=100$ TeV is derived at $99\%$ CL in a hadronic model. The lower limit of $E_\mathrm{cut,\;h}=185$ TeV at $90\%$ CL 
on the energy cutoff derived in this work from $27$h HGPS data is compatible with the result in \cite{igor} but not more constraining.\\
A detailed hadronic model for the $\gamma$-ray emission of HESS J1651-463 is presented in \cite{j1641_hadronic_model}. Within this model, runaway CR particles accelerated 
inside the young and nearby SNR G338.3-0.0 interact with a molecular cloud where $\gamma$-rays are produced via the decay of neutral pions. The SNR G338.3-0.0 itself is 
associated with the $\gamma$-ray source HESS J1640-465.\\
Given the presence of a nearby SNR and a molecular cloud, a hadronic scenario for this $\gamma$-ray source is plausible. 
Observations with instruments which have an improved sensitivity at few $10$ TeV to few $100$ TeV or a neutrino detection are necessary to confirm a Pevatron scenario.
\subsection{MGRO J1908+06}
The $\gamma$-ray source MGRO J1908+06 was discovered by Milagro \citep{milagro1908} and later confirmed by HESS \citep{hess1908}. \cite{j1908_pulsar_association} find a spatial 
coincidence between MGRO J1908+06 and the PWN of the pulsar PSR 1907.5+0602 
detected by Fermi/LAT. However, three significant emission regions are detected in the field with data from VERITAS \citep{veritas1908}. This raises the question whether the entire 
emission can originate from the PWN or an additional $\gamma$-ray source is present. As candidate for an additional $\gamma$-ray source, the SNR G40.5-0.5 is discussed in \cite{veritas1908}.\\
The HESS spectrum is confirmed by the measurement with VERITAS \citep{veritas1908}. However, the spectrum measured by ARGO-YBJ is incompatible with the 
spectra measured by HESS and VERITAS \citep{argo_1908}. HAWC detects the 
$\gamma$-ray source at an energy of $7$ TeV and derives a flux which varies within a factor of $2.5$, depending on the assumed source extension \citep{hawc_1908}. The $\gamma$-ray source is discussed 
as a possible source of high energy neutrinos \citep{neutrino_candidates}. \cite{icecube_1908} find 
a pre-trial p-value of $2.5\%$ when testing a background only hypothesis against the emission of astrophysical high energy neutrinos from MGRO J1908+06 based on data from IceCube.\\ 
It is concluded that the spectral properties of the $\gamma$-ray emission of 
MGRO J1908+06 are currently not unambiguously measured. Also the source identification is not finally resolved, although an at least partial 
association with the PWN of PSR 1907.05+0602 is likely.\\
The lower limit on the exponential cutoff of the gamma-ray spectrum derived above, $E_\mathrm{cut,\;\gamma}\approx30$ TeV, is a factor of almost $2$ more constraining than previous results 
discussed in \cite{hess1908} and \cite{veritas1908}. The lower limit $E_\mathrm{cut,\;\gamma}\approx30$ TeV is derived from HGPS data ($E_\mathrm{cut,\;\gamma}=31.1$ TeV at $90\%$ CL) 
and confirmed in an analysis of combined data from HESS, VERITAS and Milagro ($E_\mathrm{cut,\;\gamma}=29.5$ TeV at $90\%$ CL). The compatibility of the lower limits on the energy cutoff derived 
from two different datasets shows that the systematic error on the energy cutoff is not likely to be dominant in case of this source.\\
A sharp energy break is ruled out at $90\%$ CL in the energy range between $1$ TeV and $10$ TeV (see Fig \ref{j1908_break}). A search for such an energy break could, for example, be 
motivated by the sharp energy break observed around $1$ TeV in the electron spectrum measured on earth (see e.g. \cite{hess_electron}. The powerlaw index of the electron spectrum measured on earth 
changes within measured errors by $\Delta\Gamma_e=1$ at $E_\mathrm{break}=1$ TeV. 
This energy break can be interpreted as cooling break \citep{electron_break}. When electrons 
generate the $\gamma$-ray signal detected towards MGRO J1908+06 via inverse Compton scattering in the Thomson regime, a break in the spectrum of electrons by $\Delta\Gamma_e=1$ would 
translate into a change of the $\gamma$-ray spectrum index by $\Delta\Gamma=0.5$ (see e.g. \cite{hinton_hofman}). An energy break of this kind is ruled out at $90\%$ CL in the energy range 
between $1$ TeV and $10$ TeV.\\
Although the derived lower limits on the $\gamma$-ray energy cutoff are more constraining than previous lower limits and a cooling break can be ruled out in the energy range between $1$ and $10$ TeV, 
the results cannot rule out leptonic scenarios for the $\gamma$-ray emission. Deeper exposures and observations with more sensitive instruments are necessary to reveal the nature of this source.
\subsection{HESS J1852-000}
It is discussed in \cite{hess_plane} that the spectral HGPS data for this source must be treated with caution due to deviations between the main HGPS result and an independent cross check 
data analysis. HESS data for this source has been analyzed before \citep{j1852}, however, without derivation of an energy spectrum. 
No other spectral data in the TeV $\gamma$-ray energy range is publicly available. The analysis of the spectral cutoff for this source is therefore not conclusive.\\
Different scenarios for the production of $\gamma$-rays are discussed in \cite{j1852} and \cite{j1852_suzaku}. Among them is an association with the SNR Kes 78 and a nearby molecular cloud.\\
It is concluded that a hadronic source scenario, in which runaway CRs from the SNR Kes 78 illuminate a molecular cloud, can currently not be ruled out. 
The inconclusive spectral data may motivate a more detailed analysis and possibly 
further observations with current-generation Cherenkov telescopes.
\subsection{HESS J1634-472}
HESS J1634-472 is cataloged as unidentified source in the HGPS without further discussion \citep{hess_plane}. The source is also detected in a previous survey of the inner Galaxy 
with HESS \citep{first_hess_scan}. Here, the SNR G337.2+0.1 and a source of X-rays are found to be within $0.2^\circ$ of HESS J1634-472 but no association is claimed. 
\cite{j1634} discuss a detection of the source with Fermi/LAT and find that there is no counterpart pulsar in this region which is energetic enough 
to power the $\gamma$-ray source HESS J1634-472.\\
No further public TeV $\gamma$-ray data is available for this source. The interesting constraint on $E_\mathrm{cut,\;h}>108$ TeV at $90\%$ CL derived from the analysis of HGPS data 
above may motivate a dedicated re-analysis of the HESS data and possibly a deeper exposure of the source.

\section{Conclusion}
\label{conclusion_section}
Five $\gamma$-ray sources are found in the HGPS catalog for which the maximal energy of accelerated particles is, in a hadronic model, at least $100$ TeV. For at least $3$ of these sources, 
a hadronic scenario for the $\gamma$-ray emission, as result from the interaction of runaway CRs with a molecular cloud, is plausible. One of the Pevatron candidates found in the HGPS is MGRO J1908+06. 
The $\gamma$-ray spectrum of this source extends to at least $E_\mathrm{cut,\;\gamma}=30$ TeV without indication of a cutoff. The lower limit on $E_\mathrm{cut,\;\gamma}$ for this source 
is a factor of almost two more constraining than previous results. A break $\Delta\Gamma>0.5$ for this $\gamma$-ray source 
can be ruled out at $90\%$ CL in the energy range between $1$ and $10$ TeV.\\ 
The search presented in this work can only find Pevatron candidates. A conclusive identification must be performed with data from neutrino telescopes 
\citep{icecube} or $\gamma$-ray data around and above $100$ TeV. The extended air shower array HAWC \citep{hawc}, now operating for more than $3$ years, can add spectral $\gamma$-ray data 
at few $10$ TeV. An improved sensitivity and higher energies will be accessible with the upcoming $\gamma$-ray detectors CTA \citep{cta} and LHAASO \citep{lhaaso}. Finally, the development of new 
experimental techniques, such as for TAIGA/HiSCORE \citep{hiscore} will help to identify galactic Pevatrons and solve the quest of the origin of cosmic rays.
\begin{acknowledgements}
The author acknowledges financial support by the German Ministry for Education and Research (BMBF).
\end{acknowledgements}

\bibliographystyle{aa}
\bibliography{pevatron_aa_04092019}
\begin{appendix}
\section{Critical values}
\label{critical_value_appendix}
Consider, as in Sec. \ref{hypotheses}, the two nested hypotheses $H_0:\lambda=0$ and $H_1:\lambda>0$ for the single parameter $\lambda$. 
The parameter set for $\mathrm{H_0}$ is not in the interior of the parameter set of $\mathrm{H_1}$. 
The asymptotic result for the LR test statistic $-2\ln\Lambda$ under $H_0$ discussed in \cite{wilks} is therefore not applicable. A modification is, however, discussed in \cite{chernoff} 
where it is predicted that $-2\ln\Lambda$ is asymptotically distributed like an equal mixture of two random variables when $\mathrm{H_0}$ is true. 
In the considered case, one of the random variables is $\chi^2$-distributed with one degree of freedom. The other random variable has a 
Dirac delta function probability density function $\delta(0)$ (see also \cite{chernoff_read}). With this asymptotic distribution $f_\mathrm{H0}$ of the test statistic $-2\ln\Lambda$ 
under $\mathrm{H_0}$, the critical value $\Lambda_\mathrm{crit}$ of the LR test statistic can be inferred from 
$\int_{-\infty}^{\Lambda_\mathrm{crit}} dx\; f_\mathrm{H0}(x)=\mathrm{CL}$.
Let $F_\mathrm{\chi^2_1}$ denote the cumulative density functions of a $\chi^2$-distribution with one degree of freedom. Then 
$\int_{-\infty}^{\Lambda_\mathrm{crit}}dx\; f_\mathrm{H0}(x)=1/2F_\mathrm{\chi^2_1}(\Lambda_\mathrm{crit})+1/2\int_{-\infty}^{\Lambda_\mathrm{crit}}dx\;\delta(0)=\mathrm{CL}$. 
This leads to the equality $\Lambda_\mathrm{crit}=F_\mathrm{\chi^2_1}^{-1}(2\mathrm{CL}-1)$ which can be used to calculate the critical value 
$\Lambda_\mathrm{crit}$ of the LR test statistic at a given CL.\\
A similar result follows for the critical value of the asymptotic F statistic in Sec. \ref{ftest}. 
Note that the F statistic (Eq. \ref{fe0}) and the LR statistic $-2\ln\Lambda$ are related via $F=(N-3)(-2\ln\Lambda)/\mathrm{RSS(\boldsymbol{\hat{\theta}_1})}$ (see Sec. \ref{lrtest}). 
Let now $F_{1,N-3}$ be the cumulative density function for a F-distributed random variable with $1$ and $N-3$ degrees of freedom. Because the 
LR statistic is asymptotically expected to be zero in half of the tests when the null hypothesis is true, the critical value $F_\mathrm{crit}$ of the F test can again be inferred from 
$1/2F_{1,N-3}(\Lambda_\mathrm{crit})+1/2\int_{-\infty}^{\Lambda_\mathrm{crit}}dx\;\delta(0)=\mathrm{CL}$ to be $F_\mathrm{crit}=F^{-1}_{1,N-3}(2\mathrm{CL}-1)$.

\section{Existence of the lower limit}
\label{existence}
Let $F(E)$ and $F_\mathrm{crit}$ be given by Eqs. \ref{fe} and \ref{f_crit}. The lower limit on the energy cutoff is defined as solution to the equation
\begin{equation}
F(E)=F_\mathrm{crit}
	\label{fe_eq}
\end{equation}
in the interval $(0,\;E_\mathrm{cut}^\mathrm{MLE})$. Here, $E_\mathrm{cut}^\mathrm{MLE}$ is the maximum likelihood 
estimator for the energy cutoff (see Sec. \ref{confidence_interval_section}). It holds that $F_\mathrm{crit}>0$ and $F(E)$ is a continuous function. 
A solution to Eq. \ref{fe_eq} does always exist in the given interval because $F(E_\mathrm{cut}^{\mathrm{MLE}})=0$ and $F(E)\to\infty$ for $E\to0$ by definition of $F(E)$.\\
However, the solution to Eq. \ref{fe_eq} in the interval $(0,\;E_\mathrm{cut}^\mathrm{MLE})$ must not necessarily be unique when general data is considered. 
Multiple solutions can, for example, occur in case of a mismatch between the model with an exponential cutoff and the true model where the likelihood function has multiple local maxima. In this case, 
the meaning of a lower limit on the energy cutoff is questionable.\\
In any case, special care must be taken when multiple solutions to Eq. \ref{fe_eq} are found in the interval $(0,\;E_\mathrm{cut}^\mathrm{MLE})$. 
In the data analyzed in this work, all lower limits were unique when using a confidence level of $90\%$.

\section{Frequentist coverage of the lower limit}
\label{lower_limit_appendix}
Consider the interval $I$ given by Eq. \ref{lower_limit}. It is argued that $I$ has excellent frequentist coverage properties when the true energy cutoff is small such that the test power 
is large. When the true energy cutoff is large such that the test power is small, the interval $I$ overcovers and the lower limit on the energy cutoff is conservative.\\
Let in the following $F$, $F(E)$ and $F_\mathrm{crit}$ be given by Eqs. \ref{fe0}, \ref{fe} and \ref{f_crit}.
It is assumed that only one solution $E_\mathrm{cut}^{-}$ for Eq. \ref{fe_eq} in the interval $(0,E_\mathrm{cut}^\mathrm{MLE})$ exists. 
Similarly, it is also assumed that at most one solution to Eq. \ref{fe_eq} exists in the interval $(E_\mathrm{cut}^\mathrm{MLE},\infty)$. Let $E_\mathrm{cut}^{+}$ be the solution to 
Eq. \ref{fe_eq} in $(E_\mathrm{cut}^\mathrm{MLE},\infty)$, 
if it exists, and $E_\mathrm{cut}^{+}=\infty$ if no solution exists. Now consider the interval $(E_\mathrm{cut}^{-},E_\mathrm{cut}^{+})$. This interval is the acceptance interval 
$A=\{E|F(E)\leq F_\mathrm{crit}\}$ of an F-test and, as such, 
defines a confidence interval for $E_\mathrm{cut}^\mathrm{MLE}$. The confidence level of the F-test to which $A$ is the acceptance interval depends, however, on $E$. This is a result of the 
fit being constrained to $\lambda>0$.\\
When the true energy cutoff is large, $F(E)$ will need to be evaluated at large energies $E$ to obtain $A$. By definition of $F(E)$, it holds that 
$F(E)\to F$ when $E\to\infty$ and $F(E)$ is the F-test statistic for a test of a powerlaw hypothesis against a powerlaw model with exponential cutoff. 
This means that $A$ is, in the limit $E\to\infty$, the acceptance interval for an F-test constructed at confidence level CL. 
In other words, it holds that $P(E_\mathrm{true}\in A)=\mathrm{CL}$. 
Additionally, by definition of the interval $I$, it holds that $P(E_\mathrm{true}\in I)\ge P(E_\mathrm{true}\in A)$. 
Together it follows $P(E_\mathrm{true}\in I)\ge\mathrm{CL}$, i.e. the overcoverage of $I$ as a frequentist confidence interval when the true energy cutoff is large.\\
Let the true energy cutoff now be small but positive 
such that the powerlaw hypothesis is discarded at the given CL. The fit constraint $\lambda>0$ is in this case irrelevant because the best fit $\lambda$ is always large. 
By definition of $F_\mathrm{crit}$, it holds that $P(E_\mathrm{true}\in A)=2\mathrm{CL}-1$. This is because without the constraint $\lambda>0$, 
$A$ is the acceptance interval of an F-test that is constructed at confidence level $2\mathrm{CL}-1$.
Close to $E_\mathrm{cut}^\mathrm{MLE}$, $F(E)$ can be quadratically approximated, see also Fig. \ref{fefig_sym}. 
Using the nomenclature of Sec. \ref{hypotheses}, it holds that $F(E)\approx (N-3)(E-E_\mathrm{cut}^\mathrm{MLE})^2/(\mathrm{RSS}(\boldsymbol{\hat{\theta}_1})\sigma^2)$ where 
$\sigma^2$ is the inverse Fisher information of $L(\boldsymbol{\theta_1})$ evaluated at $\boldsymbol{\hat{\theta}_1}$. This means that the confidence interval $A$ will be symmetric 
around $E_\mathrm{cut}^\mathrm{MLE}$ when $E$ is small such that the solution to Eq. \ref{fe_eq} is found in the range where the quadratic approximation holds. 
It follows that $P(E_\mathrm{true}\in(0,E_\mathrm{cut}^{-}))=P(E_\mathrm{true}\in(E_\mathrm{cut}^{+},\infty))=(1-(2\mathrm{CL}-1))/2=1-\mathrm{CL}$. 
For the coverage probability, it follows that 
$P(E_\mathrm{true}\in I)=P(E_\mathrm{true}\in A)+P(E_\mathrm{true}\in (E_\mathrm{cut}^{+},\infty))=\mathrm{CL}$. This states that 
very good coverage properties of $I$ as a frequentist confidence interval are expected when the F-test power is large at the given CL. 
\begin{figure*}
	\centering
	\includegraphics[width=\hsize]{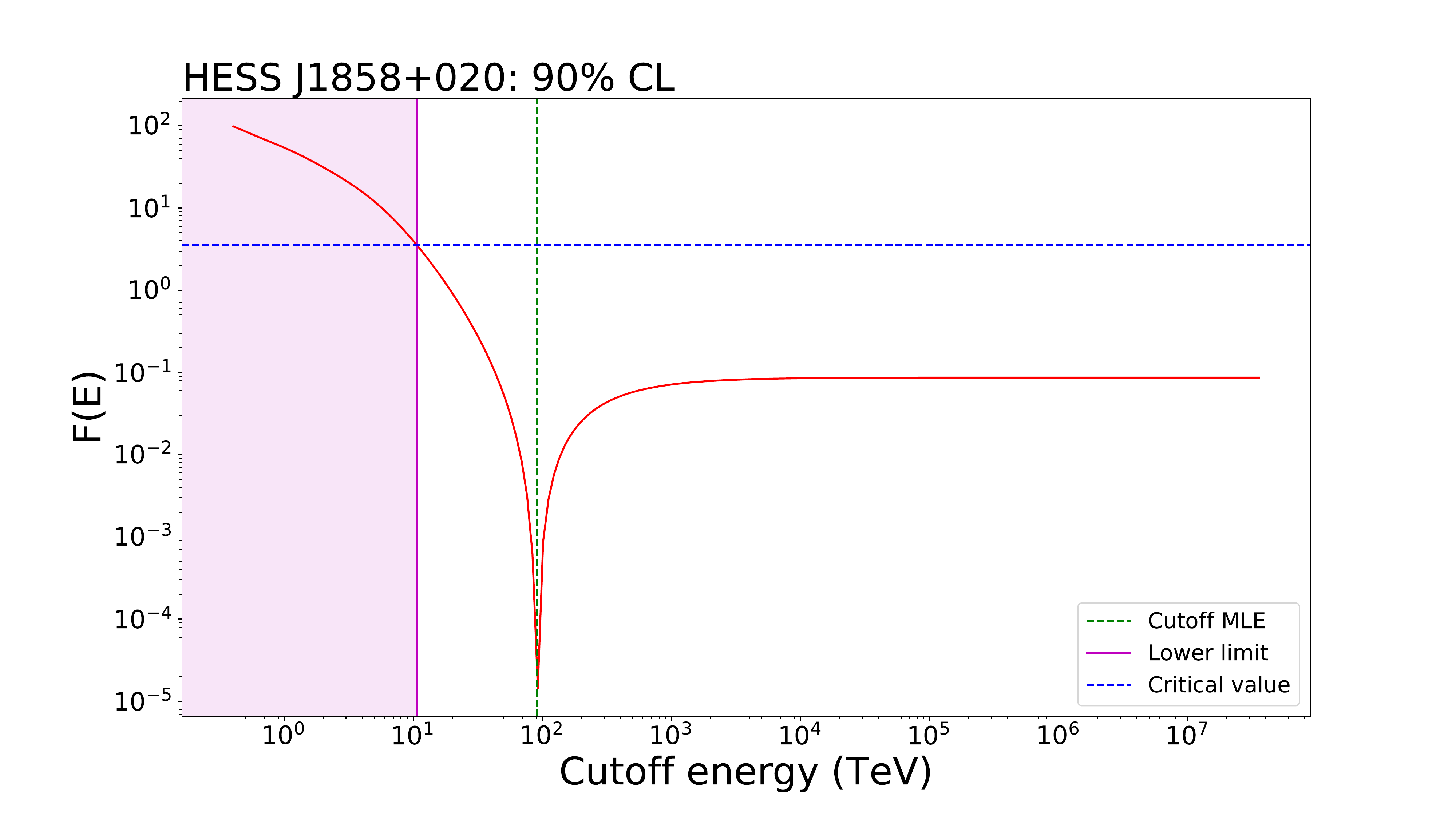}
	\caption{Example for the dependence of $F(E)$, given by Eq. \ref{fe}, on the cutoff energy $E$. The example is based on the HGPS data for the source HESS J1858+020. 
	The maximum likelihood estimate $E_\mathrm{cut,\;\gamma}^\mathrm{MLE}$ for an exponential cutoff in 
	the $\gamma$-ray spectrum (Eq. \ref{ecpl} with $\beta=1$) is indicated with a green dashed line. 
	The blue and the magenta lines indicate the critical value $F_\mathrm{crit}$ and the inferred lower limit on $E_\mathrm{cut,\;\gamma}$ at $90\%$ CL. The magenta region is the excluded 
	range of cutoff energies.}
	\label{fefig}
\end{figure*}
\begin{figure*}
	\centering
	\includegraphics[width=\hsize]{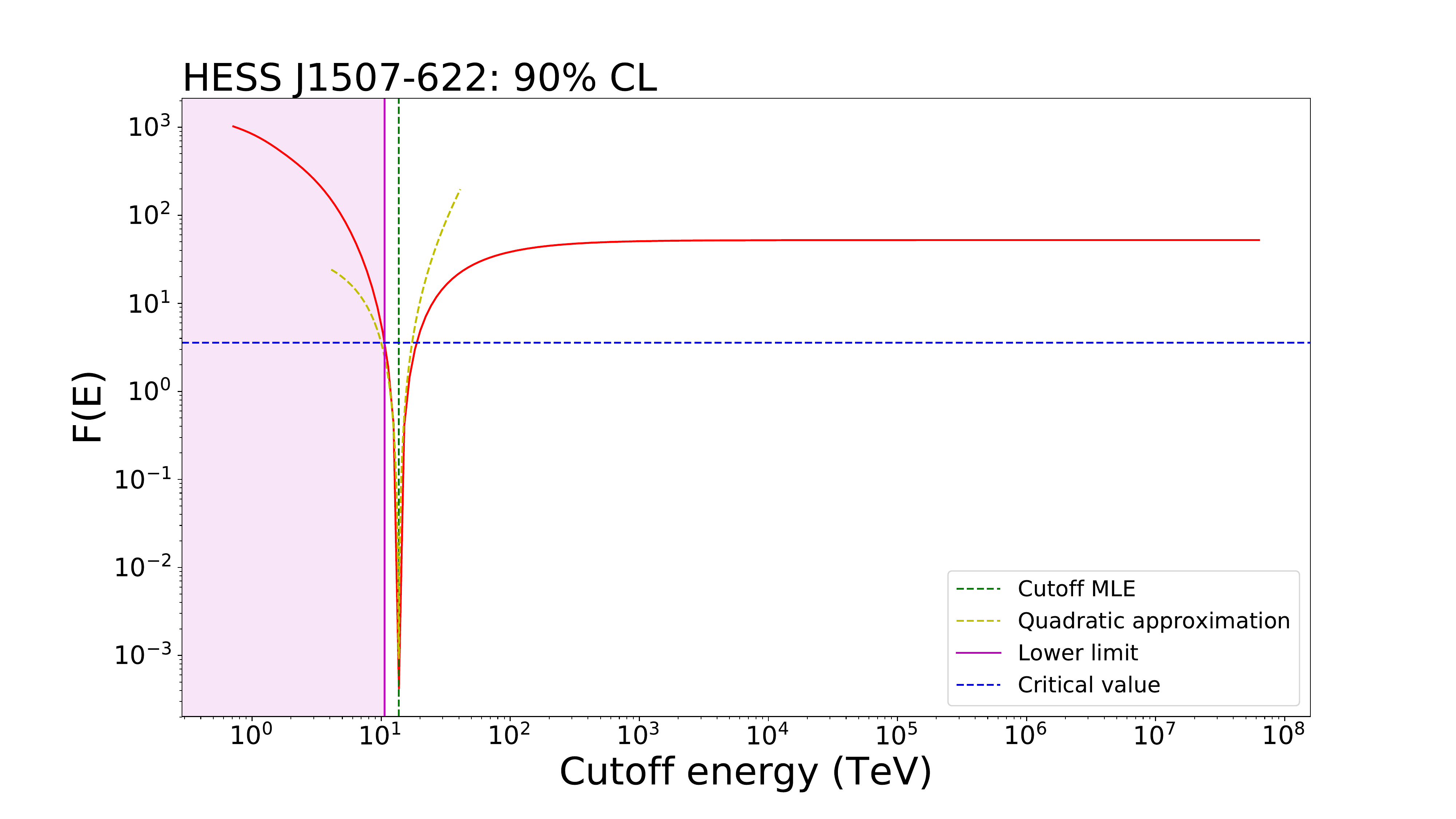}
	\caption{Same as Fig. \ref{fefig} but based on the HGPS data for the source HESS J1507-622. The powerlaw hypotheses is discarded at $90\%$ CL and 
	the acceptance interval $A=\{E|F(E)\leq F_\mathrm{crit}\}$ is now a finite interval roughly symmetric around $E_\mathrm{cut,\;\gamma}^\mathrm{MLE}$. 
	Shown in yellow is the quadratic approximation of $F(E)$ around $E_\mathrm{cut,\;\gamma}^\mathrm{MLE}$.}
	\label{fefig_sym}
\end{figure*}

\end{appendix}
\end{document}